# Diffusion Dynamics with Changing Network Composition

Raquel A. Baños [1,*], Javier Borge-Holthoefer [1], Ning Wang [2], Yamir Moreno [1,3,4] and

Sandra González-Bailón [5,★]

[1] Instituto de Biocomputación y Física de Sistemas Complejos (BIFI), Universidad de Zaragoza, Mariano Esquillor s/n, 50018 Zaragoza, Spain

[2] Oxford Internet Institute, University of Oxford, OX1 3JS Oxford, UK

[3] Departamento de Física Teórica, Universidad de Zaragoza, 50009 Zaragoza, Spain

[4] Complex Networks and Systems Lagrange Lab, Institute for Scientific Interchange, Turin, Italy

[5] Annenberg School for Communication, University of Pennsylvania, 3620 Walnut Street, PA 19104 Philadelphia, U.S.

\* Authors to whom correspondence should be addressed; E-Mails: raquel.alvarez.ba@gmail.com; sgonzalezbailon@asc.upenn.edu

**Abstract:** We analyze information diffusion using empirical data that tracks online communication around two instances of mass political mobilization, including the year that lapsed in-between the protests. We compare the global properties of the topological and dynamic networks through which communication took place as well as local changes in network composition. We show that changes in network structure underlie aggregated differences on how information diffused: an increase in network hierarchy is accompanied by a reduction in the average size of cascades. The increasing hierarchy affects not only the underlying communication topology but also the more dynamic structure of information exchange; the increase is especially noticeable amongst certain categories of nodes (or users). This suggests that the relationship between the structure of networks and their function in diffusing information is not as straightforward as some theoretical models of diffusion in networks imply.

**PACS Codes:**

# 1. Introduction

Networks offer mathematical representations of the interdependence that links decisions and behavior [1, 2, 3, 4]. Identifying the structural properties of networks can shed light into how they shape dynamic processes like the diffusion of information or cascading behavior. These dynamics underlie relevant social phenomena, for instance, political mobilizations. Recent events suggest that online networks facilitate the large-scale diffusion of protest information in relatively short time-spans [5, 6, 7, 8, 9]. However, networks are not stable structures: people discontinue their contribution to communication flows or decide to join anew. On a local level, the composition of networks changes constantly, but it remains an empirical matter to determine whether these changes generate structural shifts that impact the way in which dynamic processes unfold. This article considers that question using data on how information diffused around two political protests related to the same social movement, separated one year apart.

We analyze the structural properties of the online communication networks that formed around these two protests, which were accompanied by the diffusion of protest-related information before and after the mobilization days; we also analyze communication during the year that elapsed between these mobilizations. We compare the size of information cascades and how the composition of the network changed during the year that separates the two observations, paying special attention to how actors migrate across different regions of the communication network. We show that local changes in network structure underlie aggregated differences on how information diffused: an increase in network hierarchy goes hand in hand with a reduction in the average side of cascades. We discuss how our findings qualify models of information diffusion and, more generally, of the relationship between the structure and function of networks.

Formal models of information diffusion reveal that some network structures are more prone to generating cascades, although not always for the same reasons. The empirical prevalence of scale-free networks, where a small minority of nodes concentrates a disproportionate number of connections, spurred research on the structural and dynamic properties that derive from long-tailed degree distributions [10]. This research has suggested that the existence of a small minority of highly connected nodes, the hubs, make the networks more efficient (i.e. faster when transmitting information) and also more robust, at least to random or accidental failure [11]; but it has also qualified the role that hubs play in information diffusion: there is now consistent evidence that hubs create bottlenecks in the network, effectively acting as firewalls to global cascades [12, 13, 14]. Contrary to what one would expect in highly concentrated networks where most of the connectivity relies on a few nodes, a global diffusion of information becomes less likely in the presence of hubs; the reason is that they end up jamming the flow of information to parts of the network that are crucial for an extended chain reaction. What this research suggests, in other words, is that the structure of networks does shape the dynamics of information diffusion but not necessarily for the most intuitive reasons, given the statistical properties observed.

These theoretical models, on the other hand, often assume that the existence of a tie is enough for information to flow; in many empirical settings, however, the existence of connections does not automatically lead to activation: they are a necessary but not a sufficient condition for diffusion to happen. This is particularly the case in social networks where, in addition to capacity issues (i.e. the existence of bottlenecks in the network), nodes can decide whether to pass on information. In many empirical domains networks unfold in multiple communication layers that reflect the timing and purpose of activations, often revealing the specialization of nodes as experts or authorities in different information domains [15, 16]. The implication of these empirical observations is that there are interaction effects between the structure of static networks (i.e. the communication channels, formed by ties that are relatively stable) and the structure of dynamic networks (i.e. the more fluid layer of actual communication that emerges from tie activation). Nodes can have different positions in these domain-specific structures, and those positions might have different levels of volatility; both things can have an impact on how information flows. We consider if this is the case, and how, through the lens of empirical data.

## 2. Diffusion Events, Communication Networks, and Information Cascades

The data we analyze track online communication through Twitter during the protests that emerged in Spain in May 2011 [5] and again in May 2012 [17, 18] to celebrate the first anniversary of the protest movement. Media accounts of the events were quick to attribute to Twitter an instrumental role in the spread of calls for action and the coordination of demonstrations; this role has subsequently been spelled out by ethnographic work and interviews with protesters [19, 20]. For both events, messages using relevant protest hashtags were collected for the period of one month, spanning days preceding and following the main demonstration days (May 15 2011 and May 13 2012, respectively); we also collected activity during the intervening year. Table 1 summarizes the three data sets.

**Table 1.** Summary of data collected

|  | Protests 2011 | Protests 2012 | Intervening Period |
| --- | --- | --- | --- |
| Date range | April 25 to May 25 | April 30 to May 30 | June 1 to March 31 |
| Total number of messages | 581,749 | 1,026,292 | 555,521 |
| Total number of unique users | 85,933 | 127,930 | 115,992 |

In both years the data were collected using the streaming API, which returns a maximum of 1% of all messages published in the Twitter public timeline. The actual percentage of messages returned, however, varies depending on the filters applied and the size of the underlying population of messages of interest. The same hashtags that were used to collect messages in 2011 were again used in 2012, but the lack of control on how APIs return the sample of messages (allegedly random) means that the difference in sample sizes might only partly result from greater awareness about the protests in 2012. Although we have good reasons to believe that in 2012 the protest movement was better known by a wider public (it was no longer unexpected, as in 2011), and that this surely translated into higher levels of online activity (the movement had more time to build up a base during the passing year), the analyses that follow might overestimate the actual amount of change from year to year (if the sampling method is introducing artificial variation); however, the aggregated patterns we identify still reveal dynamics that are intrinsic to the diffusion of protest information through this online network.

Using the sampled messages, we reconstructed the communication networks for both events and the intervening year using the mentions and re-tweets (RTs) to infer links between users: if user $i$ mentions or RTs user $j$, an arc is created from $i$ to $j$; this resulted in three weighted, directed networks, one for each observation period. We also reconstructed the following/follower structure of the users sending protest messages, which resulted in unweighted, directed networks: again, if user $i$ follows user $j$, an arc is formed from $i$ to $j$. The following/follower structure was filtered so that only users active in protest communication (according to our samples) are retained. These networks, which capture the topological and dynamic structures underlying communication around the protests, are summarized in Table 2.

**Table 2.** Summary of networks

|  | 2011 | | 2011 - 2012 | | 2012 | |
| --- | --- | --- | --- | --- | --- | --- |
|  | following/er (topological) | @s (dynamic) | following/er (topological) | @s (dynamic) | following/er (topological) | @s (dynamic) |
| $N$ (# nodes) | 85,712 | 50,369 | 113,677 | 35,815 | 127,400 | 127,068 |
| $M$ (# arcs) | 6,030,459 | 135,637 | 10,191,085 | 98,709 | 7,459,518 | 522,430 |
| $<k>$ (avg degree) | 7.36 | 2.69 | 89.65 | 2.76 | 58.55 | 4.11 |
| $\max(k_{in})$ (max indegree) | 5,773 | 10,781 | 8,262 | 3,118 | 12,552 | 12,269 |
| $\max(k_{out})$ (max outdegree) | 31,798 | 245 | 37,810 | 651 | 34,892 | 658 |
| $C$ (clustering) | 0.022 | 0.002 | 0.028 | 0.015 | 0.026 | 0.013 |
| $l$ (path length) | 2.45 | 3.97 | 2.52 | 4.18 | 2.71 | 4.00 |
| $D$ (diameter) | 6 | 15 | 7 | 16 | 8 | 15 |
| $r$ (assortativity) | -0.13 | -0.07 | -0.11 | -0.09 | -0.13 | -0.08 |
| # strong components | 3,392 | 23,445 | 10,871 | 20,309 | 12,151 | 59,792 |
| $N$ giant component | 82,253 | 26,881 | 102,750 | 15,572 | 115,105 | 67,331 |
| N $2^{nd}$ component | 4 | 2 | 3 | 2 | 4 | 2 |

The diffusion curves for each event are shown in Figure 1, panel (a). The vertical axis tracks the normalized cumulative proportion of users that had sent at least one message in time $t$, as tracked by the horizontal axis, centered on the protest day. The figure shows that the acceleration rate was higher, but later, in 2011, in line with the sudden (and unexpected) explosion of the movement as protests unfolded; in 2012 the momentum started to build earlier, prior to the long-planned mass demonstrations, but at a slower pace. The degree and $k$-core distributions for the dynamic networks of mentions are shown in panels (b) and (c), respectively. Degree centrality measures the number of adjacent nodes [21]; in this case, this is the number of other users that are mentioned by a user $j$ plus those who mention $j$. The second metric, $k$-core, is a measure of cohesion; it partitions the network into groups where all nodes have the same minimum degree $k$ [22]: the higher $k$ is, the closer a user is to the denser core of the network. Panels 1(b) and 1(c) show that the 2012 network is slightly more asymmetrical, with more extreme outliers in the degree distribution; it is also significantly denser at its core, with a thicker set of nested layers of highly connected actors. Although these differences need to be interpreted cautiously given that networks differ in size, they suggest that during the passing year communication dynamics around the political movement grew more cohesive at its core, with clearer gravity centers attracting most of the activity; this core is formed by an elite of users that grew more prominent in the exchange of information.

**Figure 1. (a)** Diffusion curve of protest activity. **(b)** Degree distribution (communication network). **(c)** *k*-core distribution (communication network).

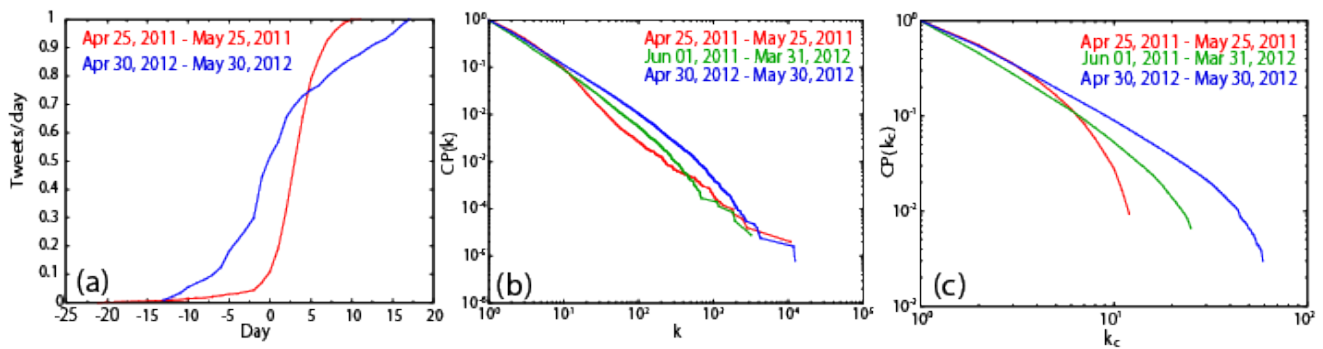

We used the size of information cascades to characterize the communication dynamics taking place in the underlying topological network. Following previous work [9, 23], we operationalize cascades assuming that activity that follows within short time periods is part of the same chain reaction. When a user sends a message at time *t*, all their followers are exposed to the information; if a little while after that, at time $t + \Delta t$, some of these followers decide to post a message as well, they are counted as part of the same cascade; so do the followers of the followers that post at $t + 2\Delta t$, and so on. The parameter $\Delta$ regulates the width of the time window used to count messages that are part of the same cascade. The final size of a cascade can then be measured as the sum of all users that send a message (we call these users 'active spreaders'), or as the total number of users that are exposed to the messages, whether or not they send a message themselves (we call these users 'listeners').

Figure 2 shows the distribution of cascade sizes both for active spreaders (panel a) and listeners (panel b), with the parameter $\Delta$ set for 1 hour (for the protest months) and for 1 day (for the intervening period). The figures show that information cascades were, for the most part, larger in 2011, although the few cases that grew extremely large reached a higher number of people in 2012 (not surprisingly, since the network is also larger). Panels (c) and (d) show the association of average cascade size with the centrality of users that started them, as measured by degree and k-core, respectively. In both cases, there is a clear association between network centrality and reach: more central users trigger, on average, cascades that activate a larger number of people; for users with similar centrality, however, cascades were larger in 2011 than in 2012 – at this level of aggregation, activity in 2012 resembles more the dormant, intermediate period separating the two protests.

**Figure 2. (a)** Cumulative cascade size probability distribution for spreaders. **(b)** Cumulative cascade size probability distribution for listeners. **(c)** Correlation between cascade size and degree centrality. **(d)** Correlation between cascades size and *k*-core.

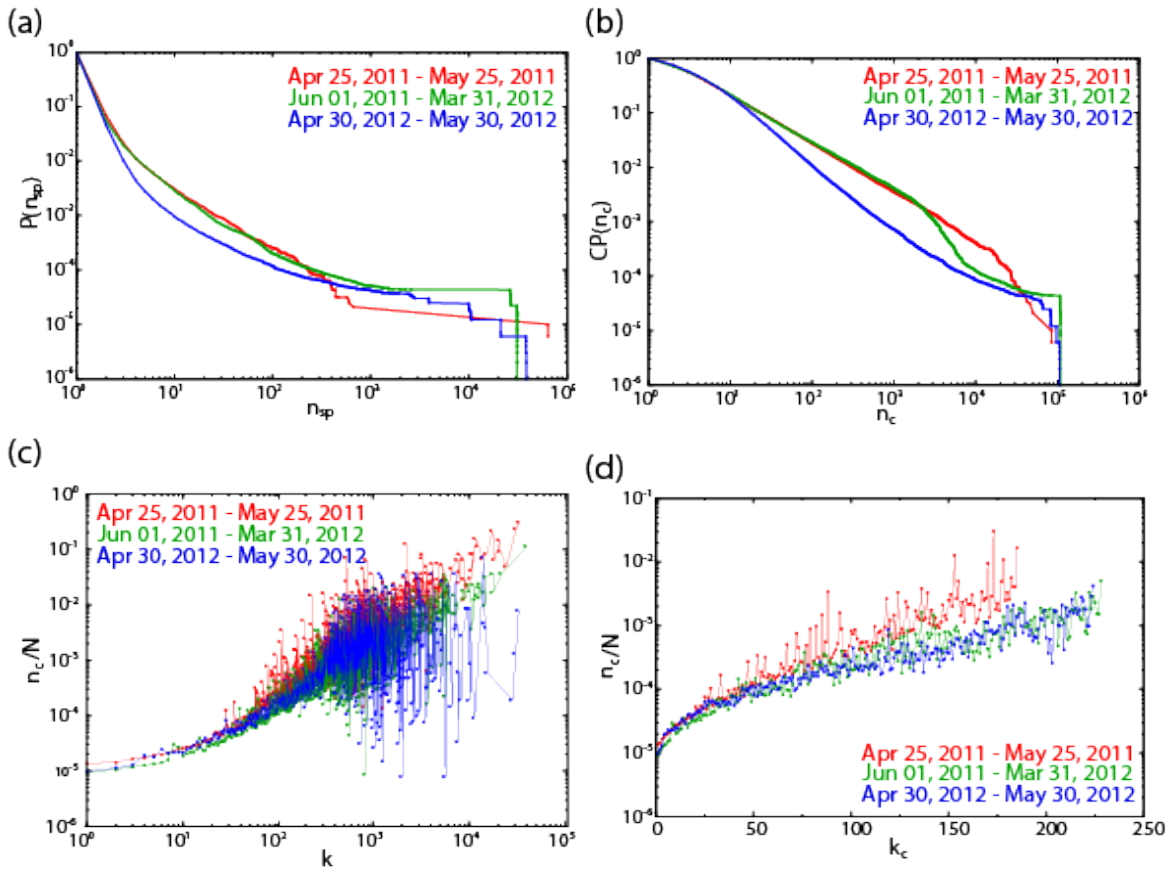

## 3. Changes in Network Composition and Information Flow

The analyses above suggest that, compared to 2011, the network in 2012 was significantly larger and denser at its core, but not necessarily as successful in terms of triggering large chains of information diffusion. During the year that separates the two observations, the network changed its composition significantly: many users that were active in 2011 disappear from the 2012 sample; even more users joined the communication exchange in 2012; and of those staying in the network, some changed their structural position. Figure 3 summarizes these differences. About 7% of the users we capture in our sample discontinued their engagement from 2011 to 2012, and about 41% joined anew in 2012; only about 8% of all users captured by the samples appear in the three observation periods. As panel (b) suggests, the network position of these users changes substantially from year to year, especially in the network of explicit protest communication.

**Figure 3. (a)** Changes in the composition of the communication network. **(b)** Correlation of centrality measures of users present in 2011 and 2012

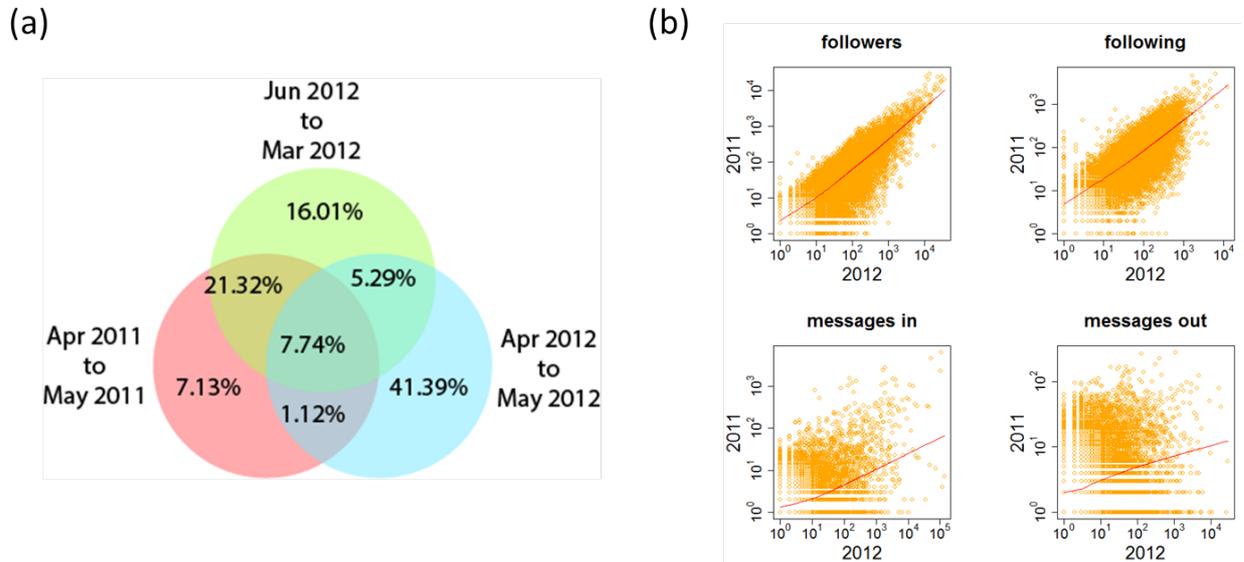

Following previous work [8], we distinguish four types of users on the basis of their position in the networks. The goal of this classification is to identify users who might not be very central in the underlying topology of the Twitter network but who have high visibility in the more dynamic, protest-specific stream of information. These users are located in quadrant 2 of the upper plots of Figure 4. We label these users 'hidden influential' to distinguish them from users who are very visible in this stream of information but also very central in the overall Twitter network – these are the users in quadrant 1, labeled 'influential' because of their larger audiences. Users in quadrants 3 and 4 are labeled 'broadcasters' and 'common users', respectively: they share a relatively lower visibility in protest communication, but those in the former category have a larger audience. What the plots reveal is that the association between network centrality and centrality in the flow of information is stronger in 2012 than in 2011, with more outliers that accumulate most of the mentions and the largest audiences. This falls in line with the increasing centrality in the degree distribution discussed above.

**Figure 4.** Distribution of users in the three observation period (upper panels) and flow of messages across categories (lower panels).

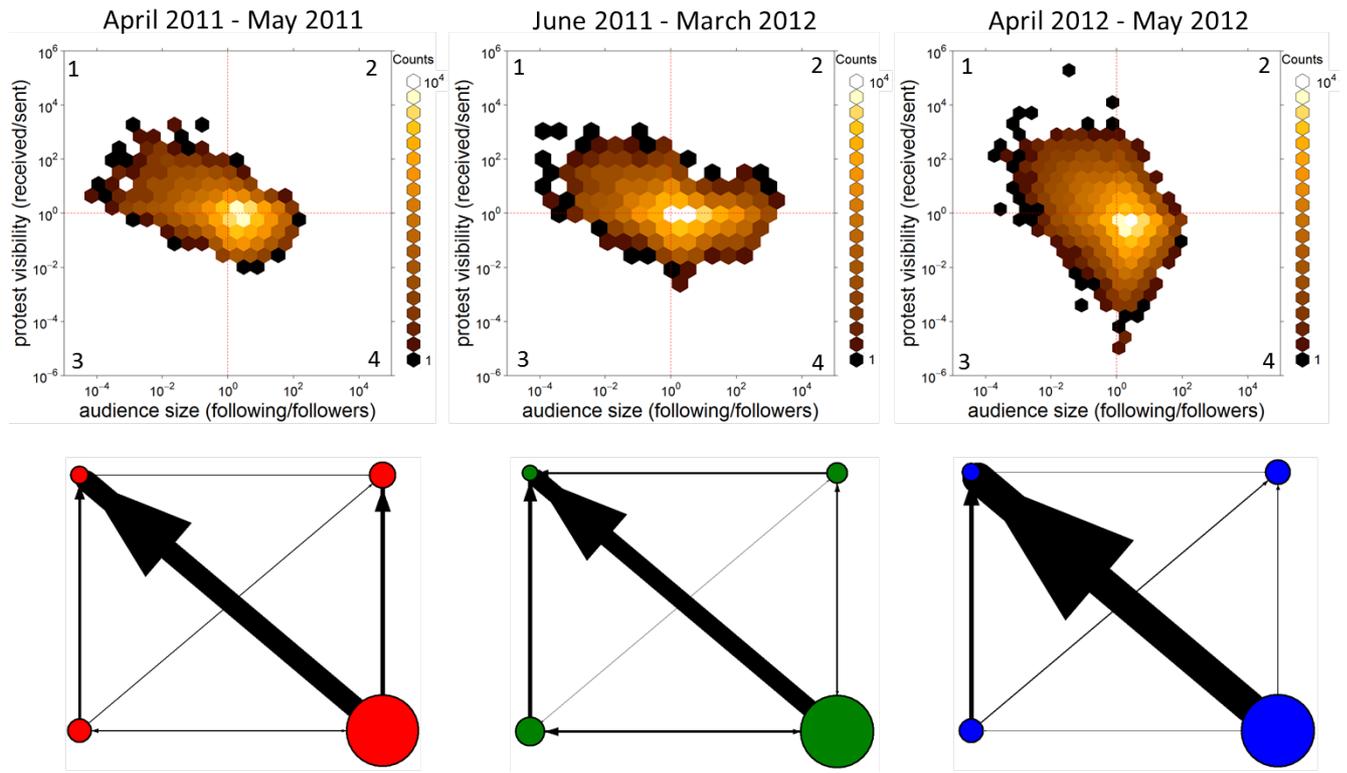

The networks in the lower panels of Figure 4 capture the dynamic aspect of the distribution of users in the four categories. These networks highlight the flow of information across groups; node sizes are proportional to the number of users classified in each group (over the total for each period), and the width of links is proportional to the number of messages coming out of each category directed to users in the other three categories (again, percentages are calculated over the total number of messages for each period). Compared to 2011, there is a visible increase in the flow of communication towards 'influential' users (quadrant 1) and a reduction in the flow towards 'hidden influential' (quadrant 2). In the intermediate period, there is a slight decentralization in the flow of communication, although a minority of users is still at the center of the exchange.

Of all the users that remain active in both years, 34% migrate across categories. As Figure 5 shows, a quarter of all migrating users go from category 4 ('common users') to category 3 ('hidden influentials'); approximately the same amount of users are downgraded in the opposite direction. The second largest stream of migrating users bridges categories 4 and 3: 16% go from being 'common' in 2011 to becoming 'broadcasters' in 2012; not many users go in the other direction (only 4%). Overall, these patterns suggest that the dynamic network of communication is more volatile than the underlying topological structure, and that path dependence is less consequential for its evolution: gaining visibility in the flow of protest related information is no guarantee that this visibility will stuck over time;

however, gaining followers, and enlarging audiences, translates into a more durable shift in the network.

**Figure 5.** Migration across categories of users from 2011 to 2012.

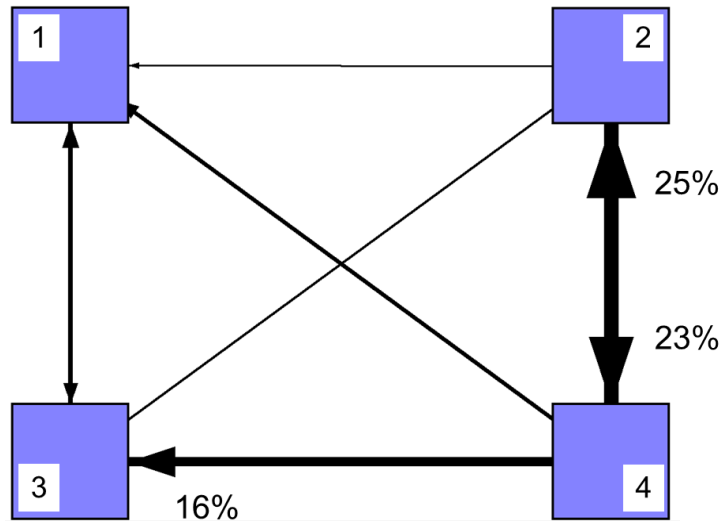

A comparison of observed and expected frequencies reveals that these differences are statistically significant (according to a resampling method [24]). Under the assumption of independence, a similar percentage of users should move across categories; however, the contingency table reveals that there is a higher number of users than random chance would make moving vertically in the network (that is, across categories 1-3 and 2-4 in Figure 5) whereas a significantly lower number of users than expected by chance move horizontally (across categories 1-2, and 3-4). This reinforces the conclusion that volatility is higher in the network of protest-related communication than in the underlying channels allowing that communication to take place. It is easier to rise and fall in the visibility around a particular topic (in this case, political protest) than in the centrality of a network that channels communication in many different domains (where politics is just one of the many). The minority of influential users in category 1 arises as the exception to this volatility: they constitute the more stable part of the network, both in terms of network centrality and visibility.

These changes in the flow of information and network composition relate to global diffusion dynamics. As suggested above, the acceleration rate of the diffusion curve is slower in 2012, and the average size of cascades is smaller; compared to the previous year, the overall network structure in 2012 was also more hierarchical and centralized. Figure 6 tries to identify the origin of this shift towards increased centralization; it displays the Gini coefficient for each category of users as a measure of inequality in the distribution of messages sent and received, for the protests in 2011 and 2012, and for the intervening period. The diagonal line acts as the benchmark of perfect equality, and a lower coefficient indicates a more equal distribution.

What Figure 6 shows is that communication activity grew more concentrated for all categories of users over time, but especially so for those classified as 'hidden influential'. After the burst of activity that accompanied the emergence of the protests in 2011, activity entered a dormant phase where a small minority of users arise as the active senders and recipients of protest information; but when activity built up again to celebrate the first anniversary in 2012, the distribution of activity remained closer to the dormant period than to the original protests. Only visibility amongst 'common users' gets closer to the 2011 levels, although it is still substantially more unequal. This means that the bottlenecks that already existed in 2011 become narrower in 2012, which partially explains why cascades grew less, on average.

**Figure 6.** Concentration of message activity by category of user.

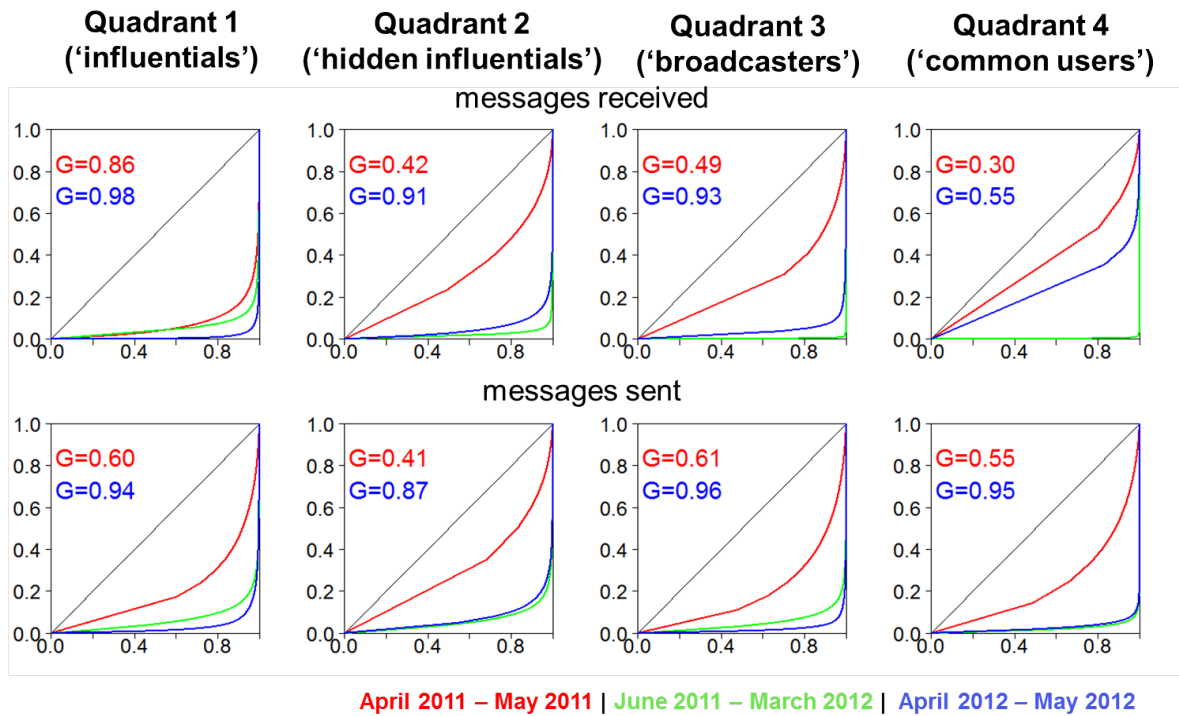

April 2011 – May 2011 | June 2011 – March 2012 | April 2012 – May 2012

The bottlenecks narrow down in several parts of the network. 'Influentials' and 'broadcasters' (who are not as visible but also have large audiences and, as a consequence, are in a position to trigger more chain reactions) are less prone to send messages in 2012 – that is, even fewer of them do. To the extent that they are the target of most messages in the network, the ratio of information received/sent becomes inevitably larger, which translates into lost information, exacerbating a phenomenon already identified in the 2011 protests [5]. Figure 6 reveals that in 2012 there were more bottlenecks both at the source and at the destination of information flows. This increasing concentration might be consistent with a preferential attachment mechanism [25], but the migration patterns identified in Figure 5 suggest that some additional mechanism, related to a differential activation of preexisting ties, might drive changes in network structure.

## 4. Discussion

The two examples of information diffusion analyzed in this paper relate to successful instances of mass political mobilization: we track communication activity that run in parallel to the build-up and explosion of massive protests. However, the mobilizations in 2012 were, by most accounts, less successful than in 2011, if only because they couldn't capitalize on the surprise factor and the passions that characterize the emergence of a political movement [19, 20]. Although it would be naïve -- and a stretch of facts – to imply that the different development of these two protests can be attributed to different uses of social media and online communication, they still offer a good case study to analyze how networks behave when diffusion follows less successful paths.

Our analyses reveal that a decrease in cascades sizes is associated with increased network centrality, both in the underlying network of connections and the more dynamic network of communication. Changes take place in the degree distributions but also in the composition of the networks: less than 10% of the users captured in 2011 re-appear in 2012, and amongst those who reappear, many change their structural position; volatility is particularly consequential in the distribution of visibility: users who were very visible in the stream of protest-related information cease being so one year later; the number of users sending the bulk of messages and therefore assigning visibility becomes also smaller, and activity more concentrated.

These changes are necessarily related to agency, that is, to the fact that nodes in this network are people who decide whether to keep on being active in the exchange of communication (political activity is, in this respect, particularly eroding). But they are also related to the interaction effects that arise from the co-evolution of networks that change at different speeds; in this case, the relatively stable structure of communication channels (the following/follower network), and the more fluid network of actual communication exchange (based on mentions). The increasing centrality in both networks creates more bottlenecks that create capacity overload; this is why cascades are smaller and the diffusion curve slower in 2012 compared to 2011. These empirical observations are consistent with previous theoretical models that highlight the detrimental role of hubs when it comes to facilitating diffusion. The changes in composition we identify, however, make networks evolve in ways that are not predicted by traditional mechanisms of preferential attachment: many nodes disappear, and many downgrade from central to peripheral positions. Networks in the real world change composition with different speed and at different rates, and this impacts their performance in ways that we are only beginning to understand.

In spite of these local changes, on the aggregate level there is a clear tendency towards increased centralization. As indicated above, we cannot be certain that these changes are reflecting a genuine trend towards more hierarchical structures or are instead a partial artefact of the sampling procedure, over which we do not have full control. However, even if we restrict the analysis to the users that appear in both observations (disregarding those who leave the network and the new additions), we still observe increasing levels of concentration; we also observe uneven migrating patterns: all users tend to focus their attention on an even narrower set of other users, but not all of them are as likely to move from one part of the network to another. These patterns of change are intriguing and not likely to result from the sampling process; they reveal local changes in information flows that run in parallel, but not necessarily in the same direction, as changes in the network structure.

Communication networks have been at the centre of diffusion studies for decades [26, 27]; but there are still many gaps in our understanding of network effects because they mediate diffusion in complex and counter-intuitive ways – especially now that technology allows networks to grow to unprecedented scale and reach [1]. Theoretical models help understand the mechanisms that shape dynamics and delimit the possibility space within which large cascades emerge; but empirical observations are still necessary to get a sense of how these dynamics operate in the real world, and how empirical instances of diffusion fit within the theoretical space of possibilities. We have provided one such empirical approach, comparing the networks and dynamics associated to the diffusion of information over the period of one year, resulting in two large-scale political mobilizations. We hope more empirical studies will follow to determine if the patterns we identify also characterize diffusion taking place in other contexts.

## Acknowledgments

R.A.B. was supported by the FPI program of the Government of Aragón, Spain. This work has been partially supported by MINECO through Grants FIS2011-25167 and FIS2012-35719; Comunidad de Aragón (Spain) through a grant to the group FENOL and by the EC FET-Proactive Projects PLEXMATH (grant 317614) and MULTIPLEX (grant 317532). S.G.B. is grateful for the financial support received from Google and the Fell Fund while still at the Oxford Internet Institute; N.W. also acknowledges support from Google.

## Conflict of Interest

The authors declare no conflict of interest.